# HTS WIGGLER CONCEPT FOR A DAMPING RING

A. Mikhailichenko, Cornell University Ithaca, NY 14853, USA
A.V. Smirnov, RadiaBeam Technologies Inc., Santa Monica, CA 90404, USA.

*Abstract*

Magnetic design proposed for a damping ring (DR) is based on second generation HTS cabling technology applied to the DC windings with a yoke and mu-metal-shimmed pole to achieve ~2T high-quality field within a 86 mm gap and 32-40 cm period. Low levels of current densities (~90-100A/mm$^2$) provide a robust, reliable operation of the wiggler at higher heat loads, up to LN$_2$ temperatures with long leads, enhanced flexibility for the cryostats and infrastructure in harsh radiation environment, and reduced failure rate compared to the baseline SC ILC DR wiggler design at very competitive cost.

## INTRODUCTION

Damping rings (DR) play a crucial role in producing the beams of sufficient quality and stability in achievement both peak and integrated luminosity in a Linear Collider. The baseline technology choice for the ILC damping ring [1], but based on the design developed for the CESR-c program [2,3]. However, the low temperature superconducting (LTS) DR wiggler is subject to failures caused by the cryogenics, power supplies, control system, and by quench. Normal-conducting alternative appears to be the most robust against long-term radiation effects. However, it would require enormously high electrical power of MW order [4]) even for small 25-mm gap TESLA DR wigglers [4,5].

Hybrid, permanent magnet ILC DR wiggler prototype having 56mm gap and ~1.7T amplitude was designed [6]. The analysis showed technical feasibility to build a prototype of such a failure-free wiggler at zero maintenance cost. Cryogenic variant of the hybrid design may sustain much higher radiation levels and also substantial heat loads, [7]. The problem is mass production: the total amount of rare-earth material required for 160-200 m wiggler (76-172 tons dependently on grade and cooling) is comparable with global year production (~130 tons in 2010 and up to 250 tons in 2015 [8]).

High Temperature Superconducting (HTS) winding made from 2nd generation wires are considered here (Figure 1). With energy cost rising and conductor cost falling, HTS magnets operating in the 20-77 K° temperature range are gaining renewed interest for the lower cost of ownership (capital and operation). Moreover, in a few low to medium field R&D applications, HTS magnets not only provided a better technical solution but also proved to be less expensive to build and test than the magnets made with conventional LTS. In addition, HTS magnets can tolerate large energy and radiation loads and can operate with a simpler cryogenic system [9].

Bismuth Strontium Calcium Copper Oxide (BSCCO) HTS wire is referred to as Generation 1 conductor [10,11]. BSCCO wire requires relatively expensive batch production process and relatively high quantities of silver (<10% of the cost). Manufacturers are transitioning to and scaling up manufacturing capacity to produce YBCO coated conductors in a semi-continuous process [12,13]. The Generation 2 technology utilizes epitaxial growth, where films deposited on a prepared structure can assume the substrate's crystal orientation. Manufacturers planning large volume price reduction in to 50% to 20% of BSCCO. That would make it competitive with copper in many large industrial applications, putting aside the cost of cooling [14].

State-of-the-art Gen.-II wire (e.g., Amperium™ [13]) presents significant leap in technological improvements to build magnets. Engineering properties exhibit good fit to the needs of ILC damping ring wiggler due to high strength and stability, hermetical solder fillets at the edges, high strength, and enhanced electrical stability, sufficient robustness, mechanical strength and bend tolerance. The 1.1-1.5cm bending radius of the wire tape is perfectly small enough to be used in 32 cm-period wiggler (though too large for some other insertion devices like short-period undulators).

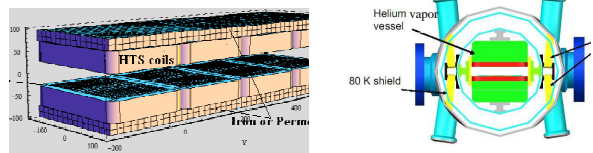

Figure 1: On the left: conceptual design of two periods of HTS wiggler. On the right: the cryostat schematics with a two phase Helium in the central chamber.

Robustness also means better sustainability to harsh radiation environment than conventional SC coils. Several HTS magnets (see, e.g., [15,16]) or insertion devices having HTS leads have been implemented [17]. Radiation-resistant dipole and quadrupole HTS magnets were developed for Rare Isotope Accelerator (RIA) [18].

## MAGNETIC DESIGN

Different modifications of the Figure 1 design to accommodate both field-gap requirements (1.97T, 86mm)

and HTS limits have been analyzed to achieve engineering (bulk coil) current density J in HTS wires, which is ~100A/mm$^2$ [19,20]. Permendur–based variant provides a viable design (see Table 1). Further optimization of Permendur-steel pole composition allows better magnetization distribution and J<100A/mm$^2$. Also Dysprosium-based variant allows reducing J down to ~90A/mm$^2$, but this variant is technically much less viable.

Table 1: Maximum magnetization, winding-averaged current density, and rare-earth mass for 1.97T field magnitude, 32mm period, 86mm gap, and 238mm pole width.

| Pole material | Magnetization, T | Averaged J, A/mm$^2$ | RE mass kg/m |
|---|---|---|---|
| Iron | 2.11 | 150 | 0 |
| NdFeB N52 | 1.63 | 106 | 132 |
| Permendur | 2.34 | 100.4 | 0 |

We estimated material-dominant wiggler cost using Table 1 for mass-produced ~2500 coils required for at least one damping ring. It turned out is ~50-75% of that compared to the baseline superferric design [1]. Important to note, that unlike the PM-hybrid variant, the wire production volume required for the two ILC damping rings is well within manufacturing capabilities today.

To compensate field roll-off in the good-field region (±10mm) we applied shimming. Despite the Iron is saturated, so each piece could be considered as an equivalent of permanent magnet with magnetization ~2T, it is possible to correct the field distribution [21,22]. The horizontal profile of the vertical magnetic field for the 86 mm gap design is given in Figure 2. For that well-shimmed variant we have dB/B=0.16% for Δx=3cm, and less than 0.0016% for Δx≤1cm (in the "good field" region). Thus the field flatness achieved is better than both the previous hybrid design (0.0045%, [6]) and the CESR-C design (0.0077% [23]) in the same "good field" region. Note, the Figure 2 means that focusing is provided in both planes and no roll-off in the good field region unlike that in the current superferric variant having defocusing in horizontal plane. That means Figure 2 variant demonstrates superior field quality.

Higher multipole field components control primarily the nonlinear properties of the wiggler transfer maps and eventually the dynamic aperture reduction caused by presence of the wigglers. There are a number of methods (e.g., generalized gradients and generalized surfaces [24,25,26]) developed for accurate field representation of insertion devices on the base of measurements or finite element simulation data. However, in our RADIA code model the material segmentation is relatively moderate and for the each segment the fields are calculated analytically. Therefore the numerical noise is considerably reduced (compared to standard OPERA-3D simulations). Therefore a conventional least square high-order approximation is applied to evaluate the multipole components. In Table 2 we summarize results for characterization of multipole components extracted from the 3D Radia field for ±10mm (horizontally) and ±6mm (vertically) areas along the 2 – period, three-fold symmetrical model.

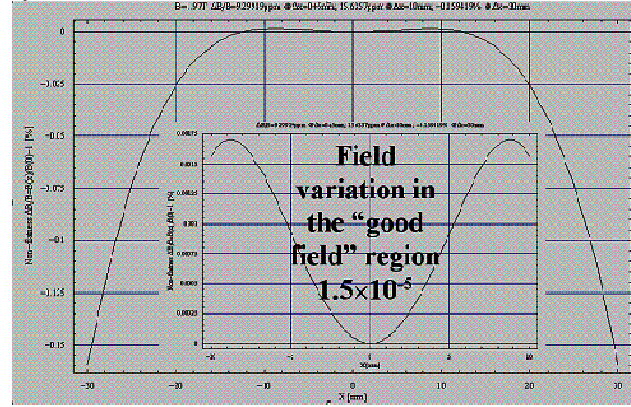

Figure 2: Field horizontal profile for 1.97T, 86mm gap wiggler having two μ-metal shims and Permendur pole. The insert: the same but for the good-field region only. The relative sextupole focusing component in the "good field" region is $k_x^2$=0.88/m$^2$.

Table 2: Multipole moments and its integrals for ±10mm (hor.) and ±6mm (vert.) areas along the 2- period short model.

|  | Qmax T/m | IQ, attoT | Smax T/m$^2$ | IS, nT/m | Octmax T/m$^3$ |
|---|---|---|---|---|---|
| Hor. | 6·10-14 | 0.7 | 0.9 | 0.1 | 1·10-9 |
| Vert | 1·10-13 | 1.7 | 144 | 0.09 | 1.2·10-8 |
|  | IOct fT/m$^2$ | Dmax T/m$^4$ | ID, μT/m$^3$ | Dodmax T/m$^5$ | IDod nT/m$^4$ |
| H | 67 | 5641. | 1.4 | 4.2·10-3 | 0.4 |
| V | 204. | 44275. | 2.8 | 0.019 | 2 |

These dominant dynamic aperture terms are determined by lattice unperturbed beta functions ($\beta_x$≈25 and $\beta_y$≈31m in our example) including wiggler linear focusing (periodic sextupole and, if any, periodic gradient) for the given 1.97T amplitude, 0.32m period, ~2 m wiggler length, 5GeV energy. It gives about $A_{xw}$≈2.2 m, $A_{yw}$≈0.1 m for a single wiggler section (0.058m and 0.01m for 80 sections correspondingly). Note this dominant (in terms of minimal dynamic aperture) term does not depend on higher order multipole components (beyond sextupole) and about the same for any wiggler (undulator) having the same period, similar field profile along the insertion device, and the same wiggler beta function(s).

Relative contribution of the multipole components in the dynamic aperture is estimated neglecting cross correlation with canonical perturbation theory [6,27]. The partial multipole dynamic apertures found are dominated by decapole component resulting from trapezoidal-like (or hat-like) horizontal profile of the vertical field. However, being compared to the dominant (i.e., minimum) partial dynamic apertures $A_{xw}$, $A_{yw}$ found above, the higher multipole contribution turns out more than an order less (for inversed value) for horizontal and more than two orders for vertical aperture. Thus the non-linear wiggler field distortions do not perturb considerably the dynamic aperture in that design.

The tune shifts are determined by periodical (linear) focusing, which includes influence of parabolic coefficient of field profile. In our model design the corresponding coefficient is only $k_x^2 \approx 0.88 \text{m}^{-2}$ (for $|\Delta x| \leq 10\text{mm}$) and the corresponding tune shifts estimated with [27] are: 0.007 and 0.0004 for horizontal and vertical planes correspondingly (compare to 0.003 and 0.04 for the hybrid variant). The shifts are much less than the incoherent tune shift and small enough to keep the operating point apart from dangerous resonances. Simultaneously the shimming suggests a useful opportunity to reduce the vertical tune shift at the expense of the horizontal one by means of increasing $k_x^2$ (i.e. moderate enhancement of horizontal focusing reduces the natural vertical focusing). Conventional acceptances, defined as the largest phase space ellipse that the wiggler and its chamber could accept, are large enough (about 1 and 0.3m·rad for the beta functions above and chamber accommodating 86 mm gap. However, positron ring admittance is more meaningful than the conventional acceptance because of injection limitation $A_x+A_y<0.09$m·rad. Assuming unperturbed ring dynamical apertures $A_{x,y}=(0.06,0.05)$m·rad we get $A_x+A_y=0.058+0.01=0.068$ m·rad for 80 wigglers using the analytics [27].

Thus presence of the wiggler in terms of dynamic aperture meets the basic ILC DR requirements as dominated primarily by its linear focusing terms that can effectively be adjusted by shimming.

## CONCLUSION

The HTS wiggler design is very competitive with a Low-Temperature Superconductor wiggler, while providing significant reduction in failure rate due to much simpler cryogenic system, sustainability to much higher magnetic fields and radiation, much wider range of temperatures, faster changes in current and "smoothed quench" specifics in HTS wires (slowed growth of the resistance).

Usage of HTS windings opens a possibility for *indirect cooling* by liquid Helium in a view of safety restriction for usage of liquid Nitrogen in confined spaces of ILC tunnels. In this case the Helium chamber not required, just tubings with liquid Helium thermally attached to the cold mass. So the cost of cryostat might be reduced substantionally. No doubdt, usage of HTS windings will be beneficiary for the ILC wiggler design.

## ACKNOWLEDGEMENTS

The authors deeply appreciate Dr. Marc Palmer from Fermilab, Mr. R.J. Rouse of Amercan Superconductor, and Arthur P. Kazanjian of SuperPower Inc. for their enthusiastic interest for this design concept.